\documentclass[]{aa} 
\usepackage{amsmath} 
\usepackage{txfonts}
\usepackage{graphicx} 
\usepackage{natbib} 
\bibpunct{(}{)}{;}{a}{}{,}

\def\chem#1{$^{#1}$}

\newcommand{\ie}{{\it i.e.}~}

\newcommand{\Ms}{\ensuremath{M_\odot}}

\newcommand{\el}[2]{\ensuremath{\rm{}^{#2}\kern-0.8pt#1}}
\newcommand{\del}[1]{\ensuremath{\delta\text{[#1/Fe]}}}
\newcommand{\kms}{km~s$^{-1}$}

\begin{document}
\title{CNO enrichment by rotating AGB stars in globular clusters}

\author{T. Decressin\inst{1,2} \and C. Charbonnel\inst{1,3} \and L. Siess\inst{4,5}
  \and A. Palacios\inst{6} \and  G. Meynet\inst{1} \and C. Georgy\inst{1}}

\offprints{T. Decressin, email: decressin@astro.uni-bonn.de}

\institute{Geneva Observatory, University of Geneva, 51, ch. des Maillettes, CH-1290
  Versoix, Switzerland 
  \and Argenlader Institut f\"ur Astronomie (AIfA),  Universit\"at Bonn, Auf dem H\"ugel 71, D-53121 Bonn,
  Germany
  \and Laboratoire d'Astrophysique de Toulouse-Tarbes,
  CNRS UMR 5572, Universit\'e de Toulouse, 14, Av. E.Belin, F-31400 Toulouse, France 
  \and Institut d'Astronomie et d'Astrophysique, Universit\'e Libre de
  Bruxelles, ULB - CP 226, B-1050 Brussels, Belgium
  \and  Centre for Stellar and Planetary Astrophysics, School of
     Mathematical Sciences, Monash University, Victoria 3800, Australia
  \and Groupe de Recherche en Astronomie et Astrophysique du Languedoc, UMR 5024, Universit\'e Montpellier II, CNRS, place Eug\`ene Bataillon, 34095 Montpellier, France}
\date{Received / Accepted}

\authorrunning{T. Decressin et al.}  

\abstract{AGB stars have long been held responsible for the important
  star-to-star variations in light elements observed in Galactic globular
  clusters.}%
{We analyse the main impacts of a first generation of rotating
  intermediate-mass stars on the chemical properties of second-generation
  globular cluster stars. The rotating models were computed without
    magnetic fields and without the effects of internal gravity waves. They
    account for the transports by meridional currents and turbulence.}%
{We computed the evolution of both standard and rotating stellar models with
  initial masses between 2.5 and 8~\Ms{} within the metallicity range
  covered by Galactic globular clusters.}%
{During central He-burning, rotational mixing transports fresh CO-rich
  material from the core towards the hydrogen-burning shell, leading to the
  production of primary \chem{14}N. In stars more massive than $M \ga
  4$\Ms, the convective envelope reaches this reservoir during the second
  dredge-up episode, resulting in a large increase in the total C+N+O
  content at the stellar surface and in the stellar wind. The corresponding
  pollution depends on the initial metallicity. At low- and
  intermediate-metallicity (i.e., [Fe/H] lower than or equal to $\sim$
  -1.2), it is at odds with the constancy of C+N+O observed among globular
  cluster low-mass stars.}%
{With the given input physics, our models suggest that massive (i.e.,
  $\ge4$~\Ms{}) rotating AGB stars have not shaped the abundance patterns
  observed in low- and intermediate-metallicity globular clusters. Our
    non-rotating models, on the other hands, do not predict surface C+N+O
    enhancements, hence are in a better position as sources of the
    chemical anomalies in globular clusters showing the constancy of the C+N+O.
    However at the moment, there is no reason to think that intermediate
    mass stars were not rotating. On the contrary there is observational
    evidence that stars in clusters have higher rotational
    velocities than in the field.}


\maketitle

\section{Introduction}

While the abundances of heavy elements (i.e., Fe-group and
$\alpha$-elements) are fairly constant from star-to-star in any
well-studied individual Galactic globular cluster (GC)\footnote{Except
  $\omega$ Cen \citep{NorrisDaCosta1995,JohnsonPilachowski2008}.}, giant
and turnoff stars present some striking anomalies in their light elements
content (Li, F, C, N, O, Na, Mg, and Al) that are not shared by their field
counterparts \citep[for a review see][]{Gratton2007}. Important is that the sum
[(C+N+O)/Fe] for individual GC stars  otherwise presenting very different
chemical features is constant to within experimental errors in all the
clusters studied so far, with the possible exception of NGC 1851 (for
references see \S5).

We now have compelling evidence that these peculiar chemical patterns were
present in the intracluster gas from which second-generation (anomalous)
low-mass stars formed, and that they resulted from the dilution of pristine
material with the hydrogen-burning products ejected by a first generation
of more massive and faster evolving GC stars \citep[for a review
see][]{PrantzosCharbonnel2007}.

Massive AGB stars that undergo efficient hot-bottom burning (HBB) during
the thermal pulse phase (TP-AGB) have been proposed as the possible GC
polluters in this so-called self-enrichment scenario
\citep{VenturaDAntona2001}.  The AGB hypothesis has been extensively
discussed in the literature, first on a qualitative basis, and more
recently with the help of custom-made standard (i.e., non-rotating) stellar
models
\citep{VenturaDAntona2001,VenturaDAntona2002,VenturaDAntona2005,VenturaDAntona2005a,VenturaDAntona2005b,VenturaDantona2008a,VenturaDAntona2008b,DenissenkovHerwig2003,KarakasLattanzio2003,Herwig2004a,Herwig2004b,FennerCampbell2004,BekkiCampbell2007}.
These studies point out several difficulties in building the observed
chemical patterns in theoretical TP-AGB models. The main problem stems from
to the competition between the third dredge-up (3DUP) that contaminates the
AGB envelope with the helium-burning ashes produced in the thermal pulse
and HBB that modifies the envelope abundances via the CNO-cycle and the
NeNa- and MgAl-chains. It is thus very difficult to obtain simultaneous O
depletion and Na enrichment in the TP-AGB envelope, while keeping the C+N+O
sum constant as required by the observations \citep[see][for more details
and references]{Charbonnel2007}.

To date, only AGB models in the mass range 5-6.5~M$_{\odot}$ have managed to
simultaneously achieve an encouraging agreement with the observed
O-depletion and the Na-enrichment \citep{VenturaDantona2008a}. These models
include the FST formulation for convection \citep{CanutoGoldman1996}, which
strongly affects the O depletion once the stars enter the TP-AGB
\citep{VenturaDAntona2005a}. Compared to the classical MLT treatment, FST
indeed leads to higher temperatures at the base of the convective envelope
(resulting in more advanced nucleosynthesis) and induces higher surface
luminosities resulting in stronger mass loss and thus fewer thermal pulses
and 3DUP events. As shown by \citet{VenturaDAntona2005a}, some O depletion
can also be obtained with the MLT prescription, but only when the free
parameter $\alpha$ is arbitrarily and strongly modified with respect to the
value calibrated on the Sun. These models are also able to produce a slight
decrease in Mg accompanied with a large increase in Al.

The Na-enrichment is more difficult to estimate and requires fine tuning of
the NeNa-cycle reaction rates. More precisely, an increase in sodium is
achieved in the most oxygen-poor ejecta of the 5-6.5~M$_{\odot}$ models
only when the maximum allowed values for the $^{22}$Ne(p,$\gamma$) rate are
adopted. In summary, and as clearly stated by \citet{VenturaDantona2008a},
the AGB scenario is viable from the nucleosynthesis point of view, provided
that only massive AGB stars of 5 to 6.5~M$_{\odot}$ contribute to the GC
self-enrichment, and under the physical assumptions described above.

However the previously quoted models have only focused on physical
uncertainties related to the TP-AGB phase, and their predictions have not
been tested in different astrophysical contexts.  In particular, the impact
of rotation in our models on the nucleosynthesis predictions for AGB
stars has never been investigated in the context of the GC self-enrichment
scenario, although rotation is often invoked to understand a wide variety
of observations \citep[e.g.][see
\S~4]{MaederMeynetARAA2000,MaederMeynet2006,ChiappiniEkstrom2008}. The
present paper addresses this question for the first time, using up-to-date
treatment for rotation-induced processes. As we shall see, the main
signature of rotation on the chemical composition of the stellar envelope
and winds of intermediate-mass stars already show up during the second
dredge-up event and cannot be erased during the TP-AGB phase.

\section{Physical input of the stellar models}
\label{sec:PI}

Although only massive AGB stars in a very narrow mass range between 5 and
6.5~\Ms{} are now suspected to play a role in the self-enrichment scenario
\citep{VenturaDantona2008a,VenturaDAntona2008b}, we computed standard and
rotating models of 2.5, 3, 4, 5, 7, and 8~\Ms{} stars with the
code STAREVOL (V2.75) \citep{SiessDufour2000,Siess2006}.  We present results
for several metallicities, namely $Z = 4 \times 10^{-3}$, $10^{-3}$, $5
\times 10^{-4}$, $10^{-4}$, and $10^{-5}$ (i.e., [Fe/H]$\simeq$ -0.66,
-1.26, -1.56, -2.26, and -3.26 respectively)\footnote{The models at
  [Fe/H]$\simeq$ -0.66 and -3.26 are actually from \citet{MaederMeynet2001}
  and \citet{MeynetMaeder2002}, respectively. To check the
  consistency of our predictions with those of the Geneva code, we 
  computed a 7~\Ms{} model at $Z=10^{-5}$ with STAREVOL and found
  excellent agreement in terms of both evolutionary and chemical
  characteristics (see Fig.~\ref{fig:cno2dup}).}. The composition is scaled
solar according to the \citet{GrevesseSauval1998} mixture and enhancement
in $\alpha$-elements ([$\alpha$/Fe] = +0.3 dex) is accounted for.  All
models were evolved up to the completion of the 2DUP.

We used the OPAL opacity tables \citep{IglesiasRogers1996} above $T>8000$~K
that account for C and O enrichments, and the \citet{FergusonAlexander2005}
data at lower temperatures. We followed the evolution of 53 chemical species
from \el{H}{1} to \el{Cl}{37} using the NACRE nuclear reaction rates
\citep{AnguloArnould1999} by default and those by
\citet{CaughlanFowler1988} otherwise \citep[see][]{SiessArnould2008}. The
treatment of convection is based on the classical mixing length formalism
with $\alpha_{\rm MLT} = 1.75$, and no convective overshoot is
included. The mass loss rates are computed with \citet{Reimers1975} formula
(with $\eta_{\rm R} = 0.5$).

For the treatment of rotation-induced processes we proceed as follows.
Solid-body rotation is assumed on the ZAMS and a typical initial surface
velocity of 300~\kms{} is chosen for all the models. (The impact of this
choice is discussed in \S~3.3 where we present models computed with initial
rotation velocities ranging from 50 to 500~\kms{}.) On the main sequence,
the evolution of the internal angular momentum profile is accounted for
with the complete formalism developed by \citet{Zahn1992} and
\citet{MaederZahn1998} (see
\citealt{PalaciosTalon2003,PalaciosCharbonnel2006,DecressinMathis2009} for
a description of the implementation in STAREVOL), which takes advection by
meridional circulation and diffusion by shear turbulence into account. The
initial solid-body profile relaxes on the main sequence on a short
timescale of a few~Myr and generates differential rotation. For our most
massive stellar models ($M\ge4$~\Ms{}), the complete formalism for angular
momentum transport is applied up to the 2DUP. For the lower mass models,
however, the complete treatment is only applied up to the end of the main
sequence, while a more crude approach is used in the more advanced
evolution phases where angular momentum evolves only through local
conservation (\ie{} only the structural changes modify the angular
momentum). This simplification is motivated by the evolutionary timescale
(i.e., the Kelvin-Helmholtz timescale on the RGB) becoming shorter than the
meridional circulation timescale.  We test this assumption by running a 5
and 7~\Ms{} with and without the full treatment of rotational mixing and
the results show little difference, validating our approximation.  In all
cases the transport of chemical species resulting from meridional
circulation and both vertical and horizontal turbulence is computed as a
diffusive process \citep{ChaboyerZahn1992} throughout the evolution.

\section{Signatures of rotation-induced mixing up to the early-AGB phase}

\subsection{Models at [Fe/H]=-1.56} 

Up to the beginning of the TP-AGB phase, the surface composition of the
standard models is only modified by the dredge-up event(s). Let us note
that only the 2.5 and 3~M$_{\odot}$ models undergo deep enough first
dredge-up so as to modify their surface abundances; however, all stars
experience to various extents the second dredge-up (hereafter 2DUP) after
central He exhaustion. As described below, in the rotating
models the effects of additional mixing become visible at the stellar
surface during the quiescent central He-burning phase.

\begin{figure*}
  {\centering
  \includegraphics[width=\textwidth]{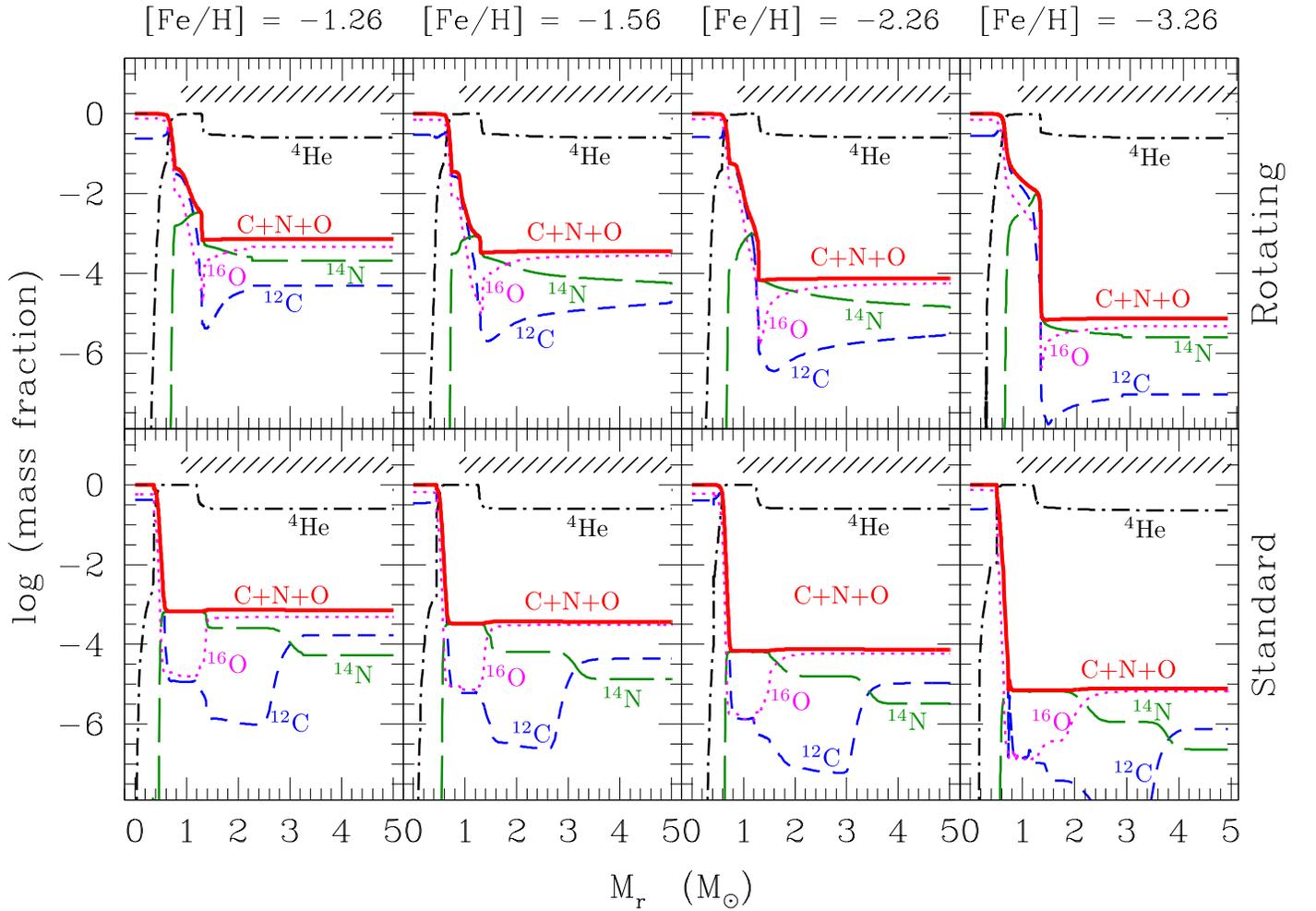}\\}
\caption{Chemical profiles (in mass fraction) at the end of central
  He-burning in rotating (top) and standard (bottom) 5~\Ms{} models at $Z =
  0.001$, 0.0005, 0.0001, and 0.00001 (left to right). The elements shown
  are \el{He}4 (dotted-dashed lines), \el{C}{12} (short-dashed lines),
  \el{N}{14} (long-dashed lines), \el{O}{16} (dotted lines), sum C+N+O
  (thick lines). The hatched area on top of each panel indicates the
  maximum extent of the convective envelope during the 2DUP.}
  \label{fig:profcno554}
\end{figure*}

Figure~\ref{fig:profcno554} shows the abundance profiles (in mass fraction)
of \el{H}1, \el{He}{4}, \el{C}{12}, \el{N}{14}, and \el{O}{16}, as well as
the total sum C+N+O, at the end of central He-burning and before the 2DUP,
for standard (lower row) and rotating (upper row) 5~\Ms{} model.

In the standard case, \chem{14}N steadily increases as one moves inwards
through the H-rich radiative layers down to the He buffer (located between
$M_r \sim 0.7$ and 1.2 \Ms) as a result of CNO processing. Further inside,
the stellar core has experienced complete He-burning and is essentially
made of \el{O}{16} and \el{C}{12}. In this 5~\Ms{} model, during the 2DUP,
the convective envelope penetrates into the He-rich buffer (down to $M_r
\simeq 0.85$~\Ms) as indicated by the hatched area. This produces an
envelope enrichment in H-burning ashes: the surface abundances of
\el{C}{12} and \el{O}{16} decrease, while that of \el{N}{14} and \el{He}4
increase. The sum C+N+O, however, remains constant, since only the H-burning
products are dredged up.

Rotation-induced mixing strongly modifies the internal chemical structure.
As shown in the upper-row panels of Fig.~\ref{fig:profcno554}, the abundance
gradients are smoothed out in the radiative envelope (i.e., the region
above the He-rich buffer up to the surface or the convective envelope)
compared to the standard case: \el{N}{14} produced in the internal
H-burning layers diffuses outwards, while the \el{C}{12} and \el{O}{16}
present in the envelope are transported inwards.  As a consequence, during 
the whole central He-burning phase, rotational mixing produces a continuous
surface increase in \el{N}{14} concomitant to a decrease in \el{C}{12} and
\el{O}{16}. At the same time, the products of central helium-burning,
namely \el{C}{12} and \el{O}{16}, also diffuse outward in the He-rich
buffer (i.e., the region where He is the dominant species, between 0.6 and
1.4~\Ms{} in Fig.~\ref{fig:profcno554}).

The transport of chemical species is mainly driven by shear turbulence
($D_\text{shear} \sim 10^8$~cm~s$^{-1}$) in the radiative envelope and in
the He-rich layer. At the interface of these two regions, i.e. at the
base of the HBS, the large mean molecular weight gradient strongly reduces
the efficiency of mixing ($D_\text{shear}\sim 10^3$~cm~s$^{-1}$), thus
preventing the transport of primary C, N, and O from the He-rich buffer to
the HBS, hence thus to the surface during central He-burning. At the same
time, hydrogen from the envelope is also transported inwards and rapidly
captured by \el{C}{12} and \el{O}{16} nuclei through CNO burning at high
temperature. This leads to the production of a peak of \emph{primary}
\el{N}{14} at the base of the hydrogen-burning shell (HBS) as seen in
Fig.~\ref{fig:profcno554}. The resulting chemical profiles at the end of
central He-burning thus differ significantly from those obtained in the
standard case where \el{N}{14} is only produced in the HBS from the
\el{C}{12} and \el{O}{16} originally present in the star and is therefore
of secondary origin \citep{MeynetMaeder2002}. During the subsequent 2DUP,
the convective envelope of the 5~M$_{\odot}$ rotating model reaches the
polluted He-rich buffer (see Fig.~\ref{fig:profcno554})
producing a large increase in \el{He}{4}. Simultaneously the primary CNO
and thus the overall metallicity increases in the envelope.

In the 2.5 and 3~\Ms{} rotating models, the convective envelope does not
reach the contaminated He-buffer during the 2DUP.  As a consequence in
these low-mass models, only the H-burning products are dredged up to the
surface: \el{N}{14} strongly increases, while \el{C}{12} and \el{O}{16}
decrease, but the sum C+N+O, as well as the total metallicity, remain
unchanged (see Table~\ref{tab:2DUP}). \el{He}4 also diffuses outward into
the radiative envelope leading to a surface \el{He}{4} enhancement by about
0.03 (in mass fraction) compared to non-rotating models.

Table~\ref{tab:2DUP} summarises the abundance variations after the 2DUP in
all our standard and rotating models at [Fe/H]=-1.56. The main signature of
rotational mixing at the surface of massive early-AGB stars ($M \ge 4~\Ms$)
is a strong increase in He-burning products, i.e., primary CNO (see also
Fig.~\ref{fig:cno2dup}). As all rotating models produce an enrichment of
CNO elements inside the He-buffer, the total increase of C+N+O at the
surface of rotating models mainly depends on the depth reached by the
convective envelope during the 2DUP.  We thus obtain a stronger variation
with increasing stellar mass.

\begin{table}[t]
  \centering
  \caption{Surface abundance variations after the completion of the 2DUP 
    with respect to the initial composition  
    ($\del{X} = \text{[X/Fe]}_\text{2DUP} - \text{[X/Fe]}_\text{init}$) 
    for the models with initial value of [Fe/H]=-1.56} 
  \label{tab:2DUP}
  \begin{tabular}{lrrrrrr}
    \hline
    \hline
    & 2.5~\Ms & 3~\Ms & 4~\Ms & 5~\Ms & 7~\Ms & 8~\Ms \\
    \hline
    & \multicolumn{6}{c}{Standard models}\\
    He       &  0.26 &  0.26 &  0.29 &  0.32 &  0.36 &  0.36 \\
    \del{C}  & -0.20 & -0.25 & -0.29 & -0.31 & -0.30 & -0.34 \\
    \del{N}  &  0.44 &  0.48 &  0.60 &  0.75 &  0.79 & 0.81 \\
    \del{O}  & -0.01 & -0.01 & -0.03 & -0.06 & -0.10 & -0.11 \\
    \del{CNO}&  0.00 &  0.00 &  0.00 &  0.00 &  0.00 & 0.00 \\
    & \multicolumn{6}{c}{Rotating models}\\
    He       &  0.29 &  0.29 &  0.32 &  0.34 &  0.35 & 0.36 \\
    \del{C}  & -1.12 & -0.86 & -0.20 &  1.44 &  2.14 & 2.16 \\
    \del{N}  &  1.23 &  1.12 &  1.13 &  1.24 &  1.19 & 1.09 \\
    \del{O}  & -0.65 & -0.36 & -0.01 &  0.18 &  1.46 & 1.55 \\
    \del{CNO}&  0.00 &  0.00 &  0.15 &  0.78 &  1.62 & 1.71 \\
    \hline
  \end{tabular}
\end{table}

\vspace{1em}

\subsection{Influence of metallicity}

\begin{figure}
  {\centering
    \includegraphics[width=0.48\textwidth]{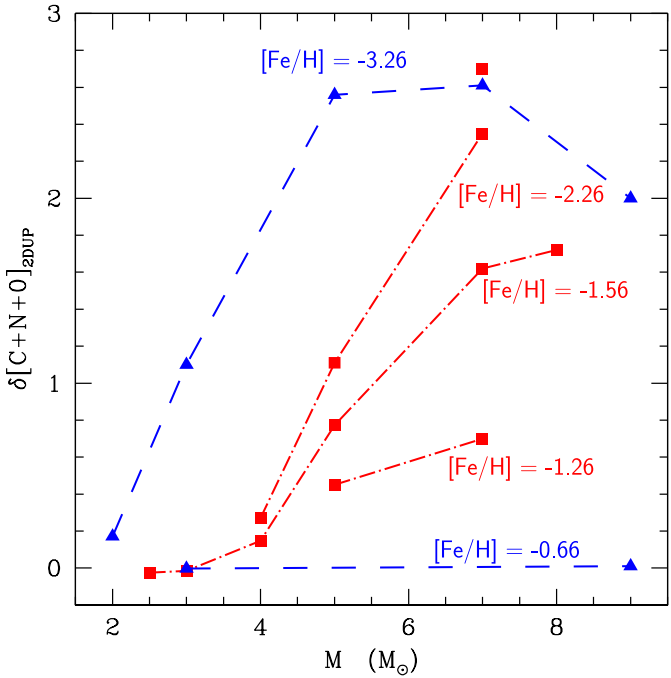}\\}
  \caption{Surface C+N+O increase index 
    ($\delta\left[\text{C+N+O}\right]_\text{2DUP} =
    [\text{C+N+O/Fe}]_\text{2DUP}-[\text{C+N+O/Fe}]_\text{init}$) at the
    end of the 2DUP for rotating stars with various initial masses and
    metallicities. Squares and triangles indicate models computed with
    STAREVOL and with the Geneva code, respectively (see text for details).
  }
  \label{fig:cno2dup}
\end{figure}

The impact of rotation on stellar properties and stellar yields is known to
depend strongly on metallicity \citep[see, e.g.,][]{MeynetMaeder2002}. The
metallicity dependence of our predictions is depicted in
Fig.~\ref{fig:profcno554} showing the internal profiles of \el{C}{12},
\el{N}{14}, and \el{O}{16} at the end of central He-burning in standard and
rotating 5~\Ms{} models at four different metallicities ($\text{[Fe/H]}
\simeq$ $-1.26$, $-1.56$, $-2.26$, and $-3.26$). As previously explained,
the C+N+O profile outside the CO core is constant in the standard models,
while it strongly increases in the He-rich layers below the HBS in the
rotating models. This C+N+O step is higher in lower metallicity stars,
which results in a stronger CNO surface enrichment after 2DUP in the most
metal-poor stars as shown in Fig.~\ref{fig:cno2dup}. The maximal depth
reached by the convective envelope for a given stellar mass hardly depend
on metallicity; e.g., it reaches 1.04 and 1.07~\Ms{} for the 7~\Ms{} at
$Z=10^{-3}$ and $Z=10^{-4}$.

At [Fe/H]=-2.26, the envelope (and thus the wind) of all the stars more
massive than $\sim$~5\Ms{} undergo a C+N+O increase by 1 to 2 orders of
magnitude, while an increase by a factor of 5 is obtained at
$\text{[Fe/H]}=-1.26$. When metallicity becomes higher than $\text{[Fe/H]}
\simeq -1$, rotation-induced mixing increases the total C+N+O by less than
a factor 2-3, which is undetectable with current observations (see
\S~5). The effect is null at $\text{[Fe/H]} = -0.66$.

\subsection{Influence of initial rotation velocity}

\begin{figure}
  \includegraphics[width=0.5\textwidth]{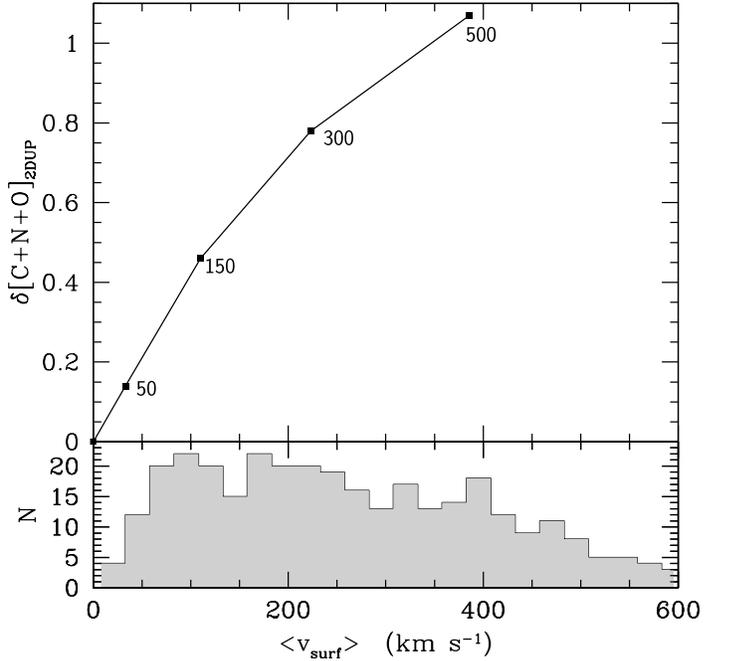}
  \caption{\textit{Top:} enrichment in C+N+O after the 2DUP completion as a
    function of the mean main sequence velocity for a 5~\Ms{},
    $Z=0.0005$ for initial surface velocities of 0, 50, 150, 300,
    500~\kms{} as indicated. \textit{Bottom:} distribution of surface
    velocities observed by \citet{MartayanFremat2006} in the young LMC
    cluster NGC~2004 for B-type stars with mass from 2 to 10~\Ms{} ($v\sin
    i$ measurements are multiplied by a factor $4/\pi$).}
  \label{fig:cnovsvsurf}
\end{figure}

We present several models for the 5~\Ms{} star at $Z=0.0005$
($\text{[Fe/H]}=-1.56$) with initial rotation velocities between 0 and
500~\kms{} as part of investigating the dependence of our results on this
parameter. As shown in Fig.~\ref{fig:cnovsvsurf}, the C+N+O enhancement
increases with increasing initial rotation. A higher initial velocity
leads to a faster spinning core and to a larger differential rotation at
the core edge during central He-burning. Since mixing stems mainly from
shear turbulence and scales as $(\partial \Omega / \partial r )^2$
\citep{TalonZahn1997}, more CNO elements are stored in the He-rich buffer
and then revealed at the surface after 2DUP. In the model with an initial
velocity of 150~\kms{}, the C+N+O rises by a factor 3, while it increases
by more than a factor 14 for an initial velocity of 500~\kms{}.

\section{Discussion on the model uncertainties}

\subsection{Mean rotation velocity}

As shown above, the total C+N+O enhancement depends on the assumed initial
rotational velocity. We note that, for an initial (i.e., ZAMS) velocity
of 300~\kms{}, the time-averaged velocity of our models on the main
sequence ranges between 220 and 256~\kms{} depending on the stellar mass
and metallicity.  What are the observational constraints on this
ingredient?

\citet{MartayanFremat2006,MartayanFloquet2007} find mean $v\sin i$ of
$161\pm20$~\kms{} and $155\pm20$~\kms{} for SMC B-type stars of 2--5~\Ms{}
(111 stars) and 5--10~\Ms{} (81 stars), respectively. Accounting in a
statistical way for the $\sin i$ effect (i.e. multiplying the averaged $v
\sin i$ by $4/\pi$ supposing a random distribution of the rotational axis),
we obtain averaged values for $v$ between 197 and 205~\kms{} for B-type
stars in the SMC.  At first glance, the main-sequence time-averaged velocity
of our models is slightly higher than the observed values for SMC
stars. However, the following points have to be considered:
\begin{enumerate}
\item The mean observed values quoted above do not account for Be-type
  stars.  Let us recall that Be stars are fast rotators that present
  emission lines originating in an outward equatorial expanding disk
  probably formed due to strong stellar rotation
  \citep{MartayanFremat2007}. Therefore Be stars belong to the upper part
  of the velocity distribution, and it is legitimate to incorporate them in
  the estimate of the averaged velocities of B-type stars. Taking them into
  account does enhance the observed average velocities. As an illustrative
  example, the mean $v \sin i$ for SMC Be stars in the mass ranges 2--5,
  5--10~\Ms{} are $277\pm34$~\kms{} (14 stars) and $297\pm25$~\kms (81
  stars), respectively. As can be seen in Fig.~\ref{fig:cnovsvsurf}, our
  assumptions on the stellar rotation velocities are thus totally
  realistic;
\item \citet{MartayanFloquet2007} \citep[see also][]{HunterLennon2008} find
  that, for both B and Be stars, the lower the metallicity, the higher the
  rotational velocities. This agrees with the finding by
  \citet{MaederGrebel1999} and \citet{WisniewskiBjorkman2006} that the
  fraction of Be stars with respect to the total number of B+Be stars
  increases when the metallicity decreases. Since the metallicities
  considered in this work are lower than that of the SMC, we may expect
  that the averaged velocities of the stars would be somewhat higher than
  the one quoted above for the SMC;
\item Last but not least, there is evidence that the rotation rates of
  stars are higher in clusters than in the field
  \citep{Keller2004,HuangGies2005,StromWolff2005,DuftonRyans2006,WolffStrom2007}.
\end{enumerate}
Thus, in view of these remarks, it does appear that our adopted rotation
velocities are probably very close to the averaged velocities of stars in
clusters for the range of masses and metallicities considered. If the
previous extrapolations are correct, the values given in
Fig.~\ref{fig:cno2dup} should be considered as lower limits.

\subsection{Treatment of rotation}

Of course, rotating models are not free from uncertainties and their
predictions should be carefully compared with well-observed features. The
physics included in the present models is the same as the one that
provides a good fit to the following observed features:
\begin{itemize}
\item the surface enrichments in nitrogen in main-sequence B-type stars
  \citep{MaederMeynet2009}, even if invoking binarity or magnetic
    fields is required to explain the whole observational pattern
    \citep{HunterBrott2008};
\item the observed number ratio of blue to red supergiants in the SMC \citep{MaederMeynet2001};
\item the variation with the metallicity of the number fraction of WR to
  O-type stars \citep{MeynetMaeder2003,MeynetMaeder2005};
\item the variation with metallicity of the number ratio of type Ibc to type II core collapse supernovae \citep{MeynetMaeder2005}.
\item the lithium abundance patterns in A-type and early F-type dwarf stars,
  as well as in their subgiant descendants
  \citep{CharbonnelTalonx1999,PalaciosTalon2003}.
\end{itemize}
Moreover, they provide a reasonable explanation for the origin of the high
N/O observed at the surface of metal-poor halo stars, as well as for
the C/O upturn
\citep{ChiappiniHirschi2006,ChiappiniEkstrom2008}. Therefore, while the
present models are not free of uncertainties, they have the nice property
of accounting for the above observed features. 

\subsection{Rotational rate of remnants}

The present models have some difficulty, however, in accounting for the
observed rotation rates of young pulsars and white dwarfs \citep[see
e.g.][]{Kawaler1988,HegerWoosley2005,SuijsLanger2008}. More precisely, they
predict too fast rotation of the stellar cores in the advanced phases. This
may be stem from different causes that could lead to additional angular
momentum loss from the central regions at different evolutionary phases:
\begin{itemize}
\item already during the nuclear lifetime;
\item at the time of the supernova explosion in the case of neutron stars
  or during the TP-AGB phase at the time of the superwind episode in the
  case of white dwarfs;
\item by the neutron stars or the white dwarfs themselves shortly after
  their formation.
\end{itemize}  
In the second and third cases mentioned above, the physics and the
predictions of the present models would not need to be revised since the
loss of angular momentum would occur after the evolutionary stages
covered by the present computations. However, if the loss of angular
momentum occurs before second dredge-up, another
mechanism should be included at earlier phases in rotating models.

The study of the s-process nucleosynthesis in low-mass TP-AGB stars could
provide some hints to the rotational evolution of the stellar core. At the
moment the properties and the behaviour of rotating TP-AGB stars are poorly
known. Calculations using a diffusive treatment for the transport of
angular momentum by \citet{LangerHeger1999}, \citet{HerwigLanger2003}, and
\citet{SiessGoriely2004} have shown that rotationally induced instabilities
provide enough mixing to trigger the 3DUPs. However, a
shear layer at the base of the convective envelope leads to an efficient
pollution of the \el{C}{13} pocket\footnote{In this framework, the
    neutrons needed for the s-process are released in the \el{C}{13} pocket
    by \el{C}{13}($\alpha,$n) reaction.} (the neutron source) by
\el{N}{14} with the result of strongly inhibiting the s-process
nucleosynthesis. This is at odds with the observations of s-stars and tends
to indicate that the modeling of rotation used in these models needs to be
improved and that angular momentum must have been removed by the time the
thermal pulses start. Therefore, if the extraction of angular momentum
occurs before core helium burning, no CNO enrichment will be expected because it
relies on shear due to a fast spinning core. On the other hand, if the
extraction occurs during the early-AGB phase, it could allow both the CNO
enrichment of the He-buffer and the s-process during the TP-AGB phase.

The mechanism frequently pointed to for removing angular momentum from
the core is the magnetic field. For instance, \citet{HegerWoosley2005} have
shown that magnetic coupling between the core and envelope can account for
the rotation rates of young neutron stars. \citet{SuijsLanger2008} also
find that magnetic torques may be required to understand the slow spin rate
of white dwarfs. However the dynamo model \citep{Spruit2002} on which the
current rotating models with magnetic fields are based has recently been
tested through hydrodynamical computations by \citet{ZahnBrun2007}, who do
not find the amplification of the magnetic field as expected from the
theory, casting some doubt on its validity. In view of the remaining
theoretical uncertainties associated with the treatment of magnetic fields,
it seems reasonable to stick to the present models whose predictions
account for a broad variety of observations as described in \S~4.2.

Finally let us note that \citet{TalonCharbonnel2008} predict that angular
momentum transport by internal gravity waves should be efficient in
intermediate-mass stars during the early-AGB phase. If these waves were the
culprit, then the predictions of the present models would be valid as far
as the CNO enrichment is concerned since it builds up earlier in the life
of the star.

\section{Consequences for the self-enrichment scenario}

\subsection{Summary of the theoretical predictions}

During central He-burning, rotational mixing efficiently transports primary
\chem{12}C and \chem{16}O outside the convective core in the H-burning
region where these elements are processed by the CNO cycle, resulting in an
important production of primary \chem{14}N. In the massive ($M \ga 4$\Ms)
rotating models, after central He-exhaustion, the convective envelope
penetrates into the layers affected by rotation-induced mixing. In contrast
to standard models, the 2DUP produces a large surface enrichment in total
C+N+O that cannot be erased by hot bottom (hydrogen) burning during the
subsequent TP-AGB evolution. Third dredge-up episodes further increase the
total C+N+O mass fraction as they bring the products of He-burning to the
surface.

As a consequence, if rotating massive AGB stars were responsible for the
abundance patterns observed in GCs, one would expect large C+N+O
differences between (O-rich and Na-poor) first-generation stars and (O-poor
and Na-rich) second-generation stars \citep[see
e.g.][]{PrantzosCharbonnel2006}. Such differences would easily amount to
1.6 dex in GCs with [Fe/H]$\sim$-1.5, and to 2.2 dex in the most metal-poor
clusters with [Fe/H]$\sim$-2.2 for the cases where the ejecta of the
first-generation would not have been diluted with pristine
material. Diluting these ejecta with 10 times more pristine material would
still give differences of 0.6 to 1.2 dex, well above the observed
dispersion (see \S~5.2). Actually, to maintain C+N+O constant would require
such a high dilution of the ejecta with the intracluster matter that the
O-Na and Mg-Al anticorrelations would be erased and, as far as abundances
are concerned, first and second generation stars would be
indistinguishable.

\subsection{Comparison with the observations}

C, N, and O abundances have been determined simultaneously in stars of
several GCs. Up to now, no significant star-to-star variation of the total
C+N+O has been detected, except in NGC 1851.

In the case of the most metal-rich GCs such as 47~Tuc
\citep{CarrettaGratton2005} and NGC~6712 \citep{YongMelendez2008}, this
C+N+O constancy is actually compatible with pollution by both standard and
rotating AGB stars since the 2DUP CNO enrichment is found to be
negligible in the most metal-rich rotating models. This is of course
without considering possible C+N+O increase due to the 3DUP of He-processed
material during the TP-AGB phase, which is beyond the scope of the present
study.

In the case of intermediate-metallicity ($\text{[Fe/H]} \sim -1.2$) GCs,
such as NGC~288 and NGC~362 \citep{Dickens1991} and M4
\citep{IvansSneden1999,SmithCunha2005}, the total C+N+O is found constant
from star to star, within observational uncertainties that amount to
$\pm$0.25--0.3 dex. At this metallicity, the most massive intermediate-mass
rotating models ($M\gtrsim5$) predict a C+N+O increase between 0.5 and 0.7
dex at odds with the observations. The only exception is NGC~1851
\citep[$\text{[Fe/H]} \sim -1.2$]{YongGrundahl2009}, where variations in
the total CNO by 0.57 $\pm$ 0.15 dex were found, which is compatible with
the CNO enrichment produced by our rotating models as shown in
Fig.~\ref{fig:cno2dup}. We note, however, that the stars observed by Yong
and collaborators in NGC~1851 are very bright objects. In view of its
position in the colour-magnitude diagram, the only star with C+N+O
exceeding the typical error bar could actually be an AGB star. In that
case, the high value of the total CNO may not have been inherited at birth
by the star, but may instead be due to nuclear processes within the star
itself. Being an AGB would also explain why this star is the only one among
the sample to exhibit some enhancement in s-process elements. We note that
in this cluster a double subgiant branch has been detected by photometric
measurements \citep{MiloneBedin2008} that can be interpreted as caused by an
age difference among cluster stars of about 1~Gyr~\footnote{The hypothesis
  of two populations with similar age and [Fe/H] differing beyond the
  errorbars have been ruled out by the spectroscopic study of RGB stars by
  \citet{YongGrundahl2009} and by the photometric study of RR Lyrae by
  \citet{Walker1998}.}. However, \citet{CassisiSalaris2008} propose to fit
the double sequence assuming two coeval cluster stellar populations with a
C+N+O difference of a factor of 2. CNO measurements in NGC~1851 unevolved
stars (as done in most of the other clusters studied so far) are thus
crucial for settling the problem.
 
Finally, the CNO predictions for the rotating most massive intermediate
mass models lead to increasing factors up to 10 (5~\Ms{}) and 100 (7~\Ms{})
at low-metallicity ($\text{[Fe/H]} = -2.2$), in clear contradiction with
the observations in the metal-poor clusters studied so far, i.e., M3 and
M13 \citep{SmithShetrone1996,Cohen2005}, NGC~6752, and
NGC~6397\citep{CarrettaGratton2005}\footnote{NGC~6397
  ($\text{[Fe/H]}\sim -1.95$) does not show a very extended O-Na
  anticorrelation, a more modest C+N+O increase is expected in that
  case.}. CNO measurements in other very metal-poor globular clusters like
M15 or M92 would be extremely valuable in this context.

\section{Conclusions}

In this paper we have investigated the effects of rotation in
low-metallicity intermediate-mass stars during their evolution up to the
completion of the 2DUP and arrival on the AGB. Our rotating stellar models
include the complete formalism developed by \citet{Zahn1992} and
\citet{MaederZahn1998} and accounts for the transport of chemicals and
angular momentum by meridional circulation and shear turbulence. With
respect to standard models, the most important change concerning the
nucleosynthesis is the large \chem{14}N production resulting from the
diffusion of protons below the HBS in a region enriched with primary C and
O during central He-burning. During the subsequent 2DUP, the convective
envelope of massive AGB stars deepens in this region, producing a large
surface increase in the total C+N+O, which irreversibly imprints the yields.

This behaviour is in sharp contrast to what is observed in low- and
intermediate-metallicity globular clusters where the sum C+N+O is constant
within the experimental errors. Our rotating models based on the
\citet{Zahn1992} formalism, which neglect the effects of magnetic fields
and internal gravity waves, thus suggesting that massive rotating AGB stars
can be discarded as potential polluters in the self-enrichment scenario in
globular clusters, unless the crowded environment prevented
intermediate-mass stars from rotating. This latest hypothesis is highly
improbable in view of the observations finding higher stellar rotation
velocities in clusters than in the field.

\begin{acknowledgements}
  We acknowledge financial support of the French Programme National de
  Physique Stellaire (PNPS) CNRS/INSU and of the Swiss National Science
  Foundation (FNS). LS is FNRS Research Associate and acknowledges
  financial support from the the Communaut\'e fran\c caise de Belgique -
  Actions de Recherche Concert\'ees. We thank E.~Carretta and A.~Bragaglia
  for enlightening discussions.
\end{acknowledgements}

\bibliographystyle{aa} 
\bibliography{BibADS}

\begin{thebibliography}{78}
\expandafter\ifx\csname natexlab\endcsname\relax\def\natexlab#1{#1}\fi

\bibitem[{{Angulo} {et~al.}(1999){Angulo}, {Arnould}, {Rayet}, {Descouvemont},
  {Baye}, {Leclercq-Willain}, {Coc}, {Barhoumi}, {Aguer}, {Rolfs}, {Kunz},
  {Hammer}, {Mayer}, {Paradellis}, {Kossionides}, {Chronidou}, {Spyrou},
  {degl'Innocenti}, {Fiorentini}, {Ricci}, {Zavatarelli}, {Providencia},
  {Wolters}, {Soares}, {Grama}, {Rahighi}, {Shotter}, \& {Lamehi
  Rachti}}]{AnguloArnould1999}
{Angulo}, C., {Arnould}, M., {Rayet}, M., {et~al.} 1999, Nuclear Physics A,
  656, 3

\bibitem[{{Bekki} {et~al.}(2007){Bekki}, {Campbell}, {Lattanzio}, \&
  {Norris}}]{BekkiCampbell2007}
{Bekki}, K., {Campbell}, S.~W., {Lattanzio}, J.~C., \& {Norris}, J.~E. 2007,
  \mnras, 267

\bibitem[{{Canuto} {et~al.}(1996){Canuto}, {Goldman}, \&
  {Mazzitelli}}]{CanutoGoldman1996}
{Canuto}, V.~M., {Goldman}, I., \& {Mazzitelli}, I. 1996, \apj, 473, 550

\bibitem[{{Carretta} {et~al.}(2005){Carretta}, {Gratton}, {Lucatello},
  {Bragaglia}, \& {Bonifacio}}]{CarrettaGratton2005}
{Carretta}, E., {Gratton}, R.~G., {Lucatello}, S., {Bragaglia}, A., \&
  {Bonifacio}, P. 2005, \aap, 433, 597

\bibitem[{{Cassisi} {et~al.}(2008){Cassisi}, {Salaris}, {Pietrinferni},
  {Piotto}, {Milone}, {Bedin}, \& {Anderson}}]{CassisiSalaris2008}
{Cassisi}, S., {Salaris}, M., {Pietrinferni}, A., {et~al.} 2008, \apjl, 672,
  L115

\bibitem[{{Caughlan} \& {Fowler}(1988)}]{CaughlanFowler1988}
{Caughlan}, G.~R. \& {Fowler}, W.~A. 1988, Atomic Data and Nuclear Data Tables,
  40, 283

\bibitem[{{Chaboyer} \& {Zahn}(1992)}]{ChaboyerZahn1992}
{Chaboyer}, B. \& {Zahn}, J.-P. 1992, \aap, 253, 173

\bibitem[{{Charbonnel}(2007)}]{Charbonnel2007}
{Charbonnel}, C. 2007, in Astronomical Society of the Pacific Conference
  Series, Vol. 378, Why Galaxies Care About AGB Stars: Their Importance as
  Actors and Probes, ed. F.~{Kerschbaum}, C.~{Charbonnel}, \& R.~F. {Wing},
  416--+

\bibitem[{{Charbonnel} \& {Talon}(1999)}]{CharbonnelTalonx1999}
{Charbonnel}, C. \& {Talon}, S. 1999, \aap, 351, 635

\bibitem[{{Chiappini} {et~al.}(2008){Chiappini}, {Ekstr{\"o}m}, {Meynet},
  {Hirschi}, {Maeder}, \& {Charbonnel}}]{ChiappiniEkstrom2008}
{Chiappini}, C., {Ekstr{\"o}m}, S., {Meynet}, G., {et~al.} 2008, \aap, 479, L9

\bibitem[{{Chiappini} {et~al.}(2006){Chiappini}, {Hirschi}, {Meynet},
  {Ekstr{\"o}m}, {Maeder}, \& {Matteucci}}]{ChiappiniHirschi2006}
{Chiappini}, C., {Hirschi}, R., {Meynet}, G., {et~al.} 2006, \aap, 449, L27

\bibitem[{{Cohen} \& {Mel{\'e}ndez}(2005)}]{Cohen2005}
{Cohen}, J.~G. \& {Mel{\'e}ndez}, J. 2005, \aj, 129, 303

\bibitem[{{Decressin} {et~al.}(2009){Decressin}, {Mathis}, {Palacios}, {Siess},
  {Talon}, {Charbonnel}, \& {Zahn}}]{DecressinMathis2009}
{Decressin}, T., {Mathis}, S., {Palacios}, A., {et~al.} 2009, \aap, 495, 271

\bibitem[{{Denissenkov} \& {Herwig}(2003)}]{DenissenkovHerwig2003}
{Denissenkov}, P.~A. \& {Herwig}, F. 2003, \apjl, 590, L99

\bibitem[{{Dickens} {et~al.}(1991){Dickens}, {Croke}, {Cannon}, \&
  {Bell}}]{Dickens1991}
{Dickens}, R.~J., {Croke}, B.~F.~W., {Cannon}, R.~D., \& {Bell}, R.~A. 1991,
  \nat, 351, 212

\bibitem[{{Dufton} {et~al.}(2006){Dufton}, {Ryans}, {Sim{\'o}n-D{\'{\i}}az},
  {Trundle}, \& {Lennon}}]{DuftonRyans2006}
{Dufton}, P.~L., {Ryans}, R.~S.~I., {Sim{\'o}n-D{\'{\i}}az}, S., {Trundle}, C.,
  \& {Lennon}, D.~J. 2006, \aap, 451, 603

\bibitem[{{Fenner} {et~al.}(2004){Fenner}, {Campbell}, {Karakas}, {Lattanzio},
  \& {Gibson}}]{FennerCampbell2004}
{Fenner}, Y., {Campbell}, S., {Karakas}, A.~I., {Lattanzio}, J.~C., \&
  {Gibson}, B.~K. 2004, \mnras, 353, 789

\bibitem[{{Ferguson} {et~al.}(2005){Ferguson}, {Alexander}, {Allard}, {Barman},
  {Bodnarik}, {Hauschildt}, {Heffner-Wong}, \&
  {Tamanai}}]{FergusonAlexander2005}
{Ferguson}, J.~W., {Alexander}, D.~R., {Allard}, F., {et~al.} 2005, \apj, 623,
  585

\bibitem[{{Gratton}(2007)}]{Gratton2007}
{Gratton}, R.~G. 2007, in Astronomical Society of the Pacific Conference
  Series, Vol. 374, From Stars to Galaxies: Building the Pieces to Build Up the
  Universe, ed. A.~{Vallenari}, R.~{Tantalo}, L.~{Portinari}, \& A.~{Moretti},
  147--+

\bibitem[{{Grevesse} \& {Sauval}(1998)}]{GrevesseSauval1998}
{Grevesse}, N. \& {Sauval}, A.~J. 1998, Space Science Reviews, 85, 161

\bibitem[{{Heger} {et~al.}(2005){Heger}, {Woosley}, \&
  {Spruit}}]{HegerWoosley2005}
{Heger}, A., {Woosley}, S.~E., \& {Spruit}, H.~C. 2005, \apj, 626, 350

\bibitem[{{Herwig}(2004{\natexlab{a}})}]{Herwig2004a}
{Herwig}, F. 2004{\natexlab{a}}, \apj, 605, 425

\bibitem[{{Herwig}(2004{\natexlab{b}})}]{Herwig2004b}
{Herwig}, F. 2004{\natexlab{b}}, \apjs, 155, 651

\bibitem[{{Herwig} {et~al.}(2003){Herwig}, {Langer}, \&
  {Lugaro}}]{HerwigLanger2003}
{Herwig}, F., {Langer}, N., \& {Lugaro}, M. 2003, \apj, 593, 1056

\bibitem[{{Huang} \& {Gies}(2005)}]{HuangGies2005}
{Huang}, W. \& {Gies}, D.~R. 2005, in Bulletin of the American Astronomical
  Society, 1273--+

\bibitem[{{Hunter} {et~al.}(2008{\natexlab{a}}){Hunter}, {Brott}, {Lennon},
  {Langer}, {Dufton}, {Trundle}, {Smartt}, {de Koter}, {Evans}, \&
  {Ryans}}]{HunterBrott2008}
{Hunter}, I., {Brott}, I., {Lennon}, D.~J., {et~al.} 2008{\natexlab{a}}, \apjl,
  676, L29

\bibitem[{{Hunter} {et~al.}(2008{\natexlab{b}}){Hunter}, {Lennon}, {Dufton},
  {Trundle}, {Sim{\'o}n-D{\'{\i}}az}, {Smartt}, {Ryans}, \&
  {Evans}}]{HunterLennon2008}
{Hunter}, I., {Lennon}, D.~J., {Dufton}, P.~L., {et~al.} 2008{\natexlab{b}},
  \aap, 479, 541

\bibitem[{{Iglesias} \& {Rogers}(1996)}]{IglesiasRogers1996}
{Iglesias}, C.~A. \& {Rogers}, F.~J. 1996, \apj, 464, 943

\bibitem[{{Ivans} {et~al.}(1999){Ivans}, {Sneden}, {Kraft}, {Suntzeff},
  {Smith}, {Langer}, \& {Fulbright}}]{IvansSneden1999}
{Ivans}, I.~I., {Sneden}, C., {Kraft}, R.~P., {et~al.} 1999, \aj, 118, 1273

\bibitem[{{Johnson} {et~al.}(2008){Johnson}, {Pilachowski}, {Simmerer}, \&
  {Schwenk}}]{JohnsonPilachowski2008}
{Johnson}, C.~I., {Pilachowski}, C.~A., {Simmerer}, J., \& {Schwenk}, D. 2008,
  \apj, 681, 1505

\bibitem[{{Karakas} \& {Lattanzio}(2003)}]{KarakasLattanzio2003}
{Karakas}, A.~I. \& {Lattanzio}, J.~C. 2003, Publications of the Astronomical
  Society of Australia, 20, 279

\bibitem[{{Kawaler}(1988)}]{Kawaler1988}
{Kawaler}, S.~D. 1988, \apj, 333, 236

\bibitem[{{Keller}(2004)}]{Keller2004}
{Keller}, S.~C. 2004, Publications of the Astronomical Society of Australia,
  21, 310

\bibitem[{{Langer} {et~al.}(1999){Langer}, {Heger}, {Wellstein}, \&
  {Herwig}}]{LangerHeger1999}
{Langer}, N., {Heger}, A., {Wellstein}, S., \& {Herwig}, F. 1999, \aap, 346,
  L37

\bibitem[{{Maeder} {et~al.}(1999){Maeder}, {Grebel}, \&
  {Mermilliod}}]{MaederGrebel1999}
{Maeder}, A., {Grebel}, E.~K., \& {Mermilliod}, J.-C. 1999, \aap, 346, 459

\bibitem[{{Maeder} \& {Meynet}(2000)}]{MaederMeynetARAA2000}
{Maeder}, A. \& {Meynet}, G. 2000, \araa, 38, 143

\bibitem[{{Maeder} \& {Meynet}(2001)}]{MaederMeynet2001}
{Maeder}, A. \& {Meynet}, G. 2001, \aap, 373, 555

\bibitem[{{Maeder} \& {Meynet}(2006)}]{MaederMeynet2006}
{Maeder}, A. \& {Meynet}, G. 2006, \aap, 448, L37

\bibitem[{{Maeder} {et~al.}(2009){Maeder}, {Meynet}, {Georgy}, \&
  {Ekstr{\"o}m}}]{MaederMeynet2009}
{Maeder}, A., {Meynet}, G., {Georgy}, C., \& {Ekstr{\"o}m}, S. 2009, in IAU
  Symposium, Vol. 259, IAU Symposium, 311--322

\bibitem[{{Maeder} \& {Zahn}(1998)}]{MaederZahn1998}
{Maeder}, A. \& {Zahn}, J.-P. 1998, \aap, 334, 1000

\bibitem[{{Martayan} {et~al.}(2007{\natexlab{a}}){Martayan}, {Floquet},
  {Hubert}, {Guti{\'e}rrez-Soto}, {Fabregat}, {Neiner}, \&
  {Mekkas}}]{MartayanFloquet2007}
{Martayan}, C., {Floquet}, M., {Hubert}, A.~M., {et~al.} 2007{\natexlab{a}},
  \aap, 472, 577

\bibitem[{{Martayan} {et~al.}(2006){Martayan}, {Fr{\'e}mat}, {Hubert},
  {Floquet}, {Zorec}, \& {Neiner}}]{MartayanFremat2006}
{Martayan}, C., {Fr{\'e}mat}, Y., {Hubert}, A.-M., {et~al.} 2006, \aap, 452,
  273

\bibitem[{{Martayan} {et~al.}(2007{\natexlab{b}}){Martayan}, {Fr{\'e}mat},
  {Hubert}, {Floquet}, {Zorec}, \& {Neiner}}]{MartayanFremat2007}
{Martayan}, C., {Fr{\'e}mat}, Y., {Hubert}, A.-M., {et~al.} 2007{\natexlab{b}},
  \aap, 462, 683

\bibitem[{{Meynet} \& {Maeder}(2002)}]{MeynetMaeder2002}
{Meynet}, G. \& {Maeder}, A. 2002, \aap, 390, 561

\bibitem[{{Meynet} \& {Maeder}(2003)}]{MeynetMaeder2003}
{Meynet}, G. \& {Maeder}, A. 2003, \aap, 404, 975

\bibitem[{{Meynet} \& {Maeder}(2005)}]{MeynetMaeder2005}
{Meynet}, G. \& {Maeder}, A. 2005, \aap, 429, 581

\bibitem[{{Milone} {et~al.}(2008){Milone}, {Bedin}, {Piotto}, {Anderson},
  {King}, {Sarajedini}, {Dotter}, {Chaboyer}, {Mar{\'{\i}}n-Franch},
  {Majewski}, {Aparicio}, {Hempel}, {Paust}, {Reid}, {Rosenberg}, \&
  {Siegel}}]{MiloneBedin2008}
{Milone}, A.~P., {Bedin}, L.~R., {Piotto}, G., {et~al.} 2008, \apj, 673, 241

\bibitem[{{Norris} \& {Da Costa}(1995)}]{NorrisDaCosta1995}
{Norris}, J.~E. \& {Da Costa}, G.~S. 1995, \apj, 447, 680

\bibitem[{{Palacios} {et~al.}(2006){Palacios}, {Charbonnel}, {Talon}, \&
  {Siess}}]{PalaciosCharbonnel2006}
{Palacios}, A., {Charbonnel}, C., {Talon}, S., \& {Siess}, L. 2006, \aap, 453,
  261

\bibitem[{{Palacios} {et~al.}(2003){Palacios}, {Talon}, {Charbonnel}, \&
  {Forestini}}]{PalaciosTalon2003}
{Palacios}, A., {Talon}, S., {Charbonnel}, C., \& {Forestini}, M. 2003, \aap,
  399, 603

\bibitem[{{Prantzos} \& {Charbonnel}(2006)}]{PrantzosCharbonnel2006}
{Prantzos}, N. \& {Charbonnel}, C. 2006, \aap, 458, 135

\bibitem[{{Prantzos} {et~al.}(2007){Prantzos}, {Charbonnel}, \&
  {Iliadis}}]{PrantzosCharbonnel2007}
{Prantzos}, N., {Charbonnel}, C., \& {Iliadis}, C. 2007, \aap, 470, 179

\bibitem[{{Reimers}(1975)}]{Reimers1975}
{Reimers}, D. 1975, {Circumstellar envelopes and mass loss of red giant stars}
  (Problems in stellar atmospheres and envelopes.), 229--256

\bibitem[{{Siess}(2006)}]{Siess2006}
{Siess}, L. 2006, \aap, 448, 717

\bibitem[{{Siess} \& {Arnould}(2008)}]{SiessArnould2008}
{Siess}, L. \& {Arnould}, M. 2008, \aap, 489, 395

\bibitem[{{Siess} {et~al.}(2000){Siess}, {Dufour}, \&
  {Forestini}}]{SiessDufour2000}
{Siess}, L., {Dufour}, E., \& {Forestini}, M. 2000, \aap, 358, 593

\bibitem[{{Siess} {et~al.}(2004){Siess}, {Goriely}, \&
  {Langer}}]{SiessGoriely2004}
{Siess}, L., {Goriely}, S., \& {Langer}, N. 2004, \aap, 415, 1089

\bibitem[{{Smith} {et~al.}(1996){Smith}, {Shetrone}, {Bell}, {Churchill}, \&
  {Briley}}]{SmithShetrone1996}
{Smith}, G.~H., {Shetrone}, M.~D., {Bell}, R.~A., {Churchill}, C.~W., \&
  {Briley}, M.~M. 1996, \aj, 112, 1511

\bibitem[{{Smith} {et~al.}(2005){Smith}, {Cunha}, {Ivans}, {Lattanzio},
  {Campbell}, \& {Hinkle}}]{SmithCunha2005}
{Smith}, V.~V., {Cunha}, K., {Ivans}, I.~I., {et~al.} 2005, \apj, 633, 392

\bibitem[{{Spruit}(2002)}]{Spruit2002}
{Spruit}, H.~C. 2002, \aap, 381, 923

\bibitem[{{Strom} {et~al.}(2005){Strom}, {Wolff}, \& {Dror}}]{StromWolff2005}
{Strom}, S.~E., {Wolff}, S.~C., \& {Dror}, D.~H.~A. 2005, \aj, 129, 809

\bibitem[{{Suijs} {et~al.}(2008){Suijs}, {Langer}, {Poelarends}, {Yoon},
  {Heger}, \& {Herwig}}]{SuijsLanger2008}
{Suijs}, M.~P.~L., {Langer}, N., {Poelarends}, A.-J., {et~al.} 2008, \aap, 481,
  L87

\bibitem[{{Talon} \& {Charbonnel}(2008)}]{TalonCharbonnel2008}
{Talon}, S. \& {Charbonnel}, C. 2008, \aap, 482, 597

\bibitem[{{Talon} \& {Zahn}(1997)}]{TalonZahn1997}
{Talon}, S. \& {Zahn}, J.-P. 1997, \aap, 317, 749

\bibitem[{{Ventura} \& {D'Antona}(2005{\natexlab{a}})}]{VenturaDAntona2005}
{Ventura}, P. \& {D'Antona}, F. 2005{\natexlab{a}}, \aap, 431, 279

\bibitem[{{Ventura} \& {D'Antona}(2005{\natexlab{b}})}]{VenturaDAntona2005a}
{Ventura}, P. \& {D'Antona}, F. 2005{\natexlab{b}}, \aap, 431, 279

\bibitem[{{Ventura} \& {D'Antona}(2005{\natexlab{c}})}]{VenturaDAntona2005b}
{Ventura}, P. \& {D'Antona}, F. 2005{\natexlab{c}}, \aap, 439, 1075

\bibitem[{{Ventura} \& {D'Antona}(2008{\natexlab{a}})}]{VenturaDantona2008a}
{Ventura}, P. \& {D'Antona}, F. 2008{\natexlab{a}}, \aap, 479, 805

\bibitem[{{Ventura} \& {D'Antona}(2008{\natexlab{b}})}]{VenturaDAntona2008b}
{Ventura}, P. \& {D'Antona}, F. 2008{\natexlab{b}}, \mnras, 385, 2034

\bibitem[{{Ventura} {et~al.}(2002){Ventura}, {D'Antona}, \&
  {Mazzitelli}}]{VenturaDAntona2002}
{Ventura}, P., {D'Antona}, F., \& {Mazzitelli}, I. 2002, \aap, 393, 215

\bibitem[{{Ventura} {et~al.}(2001){Ventura}, {D'Antona}, {Mazzitelli}, \&
  {Gratton}}]{VenturaDAntona2001}
{Ventura}, P., {D'Antona}, F., {Mazzitelli}, I., \& {Gratton}, R. 2001, \apjl,
  550, L65

\bibitem[{{Walker}(1998)}]{Walker1998}
{Walker}, A.~R. 1998, \aj, 116, 220

\bibitem[{{Wisniewski} \& {Bjorkman}(2006)}]{WisniewskiBjorkman2006}
{Wisniewski}, J.~P. \& {Bjorkman}, K.~S. 2006, \apj, 652, 458

\bibitem[{{Wolff} {et~al.}(2007){Wolff}, {Strom}, {Dror}, \&
  {Venn}}]{WolffStrom2007}
{Wolff}, S.~C., {Strom}, S.~E., {Dror}, D., \& {Venn}, K. 2007, \aj, 133, 1092

\bibitem[{{Yong} {et~al.}(2009){Yong}, {Grundahl}, {D'Antona}, {Karakas},
  {Lattanzio}, \& {Norris}}]{YongGrundahl2009}
{Yong}, D., {Grundahl}, F., {D'Antona}, F., {et~al.} 2009, \apjl, 695, L62

\bibitem[{{Yong} {et~al.}(2008){Yong}, {Mel{\'e}ndez}, {Cunha}, {Karakas},
  {Norris}, \& {Smith}}]{YongMelendez2008}
{Yong}, D., {Mel{\'e}ndez}, J., {Cunha}, K., {et~al.} 2008, \apj, 689, 1020

\bibitem[{{Zahn}(1992)}]{Zahn1992}
{Zahn}, J.-P. 1992, \aap, 265, 115

\bibitem[{{Zahn} {et~al.}(2007){Zahn}, {Brun}, \& {Mathis}}]{ZahnBrun2007}
{Zahn}, J.-P., {Brun}, A.~S., \& {Mathis}, S. 2007, \aap, 474, 145

\end{thebibliography}

\end{document}